\documentclass[nofootinbib,aps,amsfonts,superscriptaddress]{revtex4}
\usepackage{graphicx,amsmath,amssymb,bm}
\usepackage{amsfonts,amssymb,amscd,amsmath}
\usepackage{graphicx}

\usepackage{epsfig}
\usepackage{wrapfig,rotating}
\usepackage{verbatim}

\def\beq{\begin{equation}}
\def\eeq{\end{equation}}
\def\bea{\begin{eqnarray}}
\def\eea{\end{eqnarray}}

\def\fig#1{{Fig.~\ref{#1}}}

\newcommand{\as}{\alpha_s}
\newcommand{\Lb}{\left(}
\newcommand{\Rb}{\right)}

\newcommand{\p}{I\!\!P}

\def\pom{{I\!\!P}}

\begin{document}

\title{Hadron production at the LHC: Any indication of new phenomena }

\author{Eugene Levin}
\affiliation{Department of Particle Physics, Tel Aviv University, Tel Aviv 69978, Israel}
\affiliation{Departamento de F\'\i sica, Universidad T\'ecnica Federico Santa Mar\'\i a, Avda. Espa\~na 1680,
Casilla 110-V,  Valparaiso, Chile } 
\author{Amir H. Rezaeian}
\affiliation{Departamento de F\'\i sica, Universidad T\'ecnica Federico Santa Mar\'\i a, Avda. Espa\~na 1680,
Casilla 110-V, Valparaiso, Chile } 
\date{\today}

\begin{abstract}
We confront soft Pomeron and gluon saturation models with the first
LHC data on inclusive hadron production. We claim that while the first
type of models are not able to describe some part of the LHC data, the
Colour-Glass-Condensate (gluon saturation) approach gives an adequate
description of the data.  Here, we compare our published predictions
with the recently available 7 TeV data. We firmly believe that if
further experimental measurements confirm that the gluon saturation
works, it will be a major discovery.
\end{abstract}

\maketitle
%\date{\today}

%%%%%%%%%%%%%%%%%%%%%%%%%%%%%%%%%%%%%%%%%%%%
%% MAINMATTER
%%%%%%%%%%%%%%%%%%%%%%%%%%%%%%%%%%%%%%%%%%%%

%%%%%%%%%%%%%%%%%%%%%%%%%%%%%%%%%%%%%%
\section{Introduction}
%%%%%%%%%%%%%%%%%%%%%%%%%%%%%%%%%%%%%%

As it was expected the first data from the LHC are on soft
interaction at high energies such as the total and diffractive
cross-sections, the hadron inclusive production and so on. Therefore,
it is a proper time to review our understanding of these processes.

For four decades, the main tool for description of soft interaction has
been the approach based on Reggeons and Pomerons and their
interactions.  The first good news is
that actually for the first time in these four decades we obtain
Pomeron from theory. Today Pomeron is not a plausible assumption
as it was in the 70s, it is not the object of successful high energy
phenomenology but it comes out naturally from the first theory of strong
interaction: N=4 Super Yang Mills (N=4 SYM) \cite{BST}. N=4 SYM
together with AdS/CFT correspondence allows us to study 
the regime of the strong coupling constant \cite{AdS-CFT}. For the
first time we have a theory which leads to the main ingredients of the
high-energy phenomenology such as Pomerons and Reggeons, in the
limit of strong coupling.  On the other hand, N=4 SYM with small
coupling leads to normal QCD like physics \cite{POST,BFKL4} with OPE  and linear equations for DIS as well as the BFKL equation for the high energy amplitude.

First, we recall that  N=4 SYM has a simple solution  for the following set of couplings:
\beq \label{I2}
g_s\,\,=\,\,\frac{g^2_{YM}}{4 \pi}\,\,=\,\,\alpha_{YM}\,\,
\,=\,\,\frac{\lambda}{4 \pi N_c};\,\,\,\,\,\,\,R\,\,=\,\, \alpha'^{\frac{1}{2}}\,\lambda^{\frac{1}{4}};\,\,\,\,\,\,\,\,g_s\,\,\ll\,\,1;\,\,\,\,\mbox{but}\,\,\,\,\,\lambda\,\,\gg\,\,1,
\eeq
where $R$ is the radius in  $AdS_5$- metric, $N_c$ denotes the number of colours and $\alpha'$ is the slope of the Reggeons ($\alpha' \approx 1\,\text{GeV}^{-2}$) which is  intimately related to the string tension in string theory.
%\beq \label{I3}
%d s^2\,\,=\,\,\frac{R^2}{z^2}\,\Lb \,d z^2\,\,+\,\,\sum^d_{i=1}  d x^2_i \Rb\,=
%\,\frac{R^2}{z^2}\,\Lb \,d z^2\,+\,dx_\mu dx^\mu \Rb
%\eeq
%with $\mu = 0,1,2,3$. 
The Pomeron which appears in N=4 SYM \cite{BST} at large $\lambda$, has a intercept and a slope of the trajectory  that are equal to
\beq\label{I1}
\alpha_{\p}\Lb 0 \Rb\,\,\,=\,\,\,2\,\,-\,\,\frac{2}{\sqrt{\lambda}}\,\,\,\,\,\,\,\,\,\,\,\,\,\,\,
\alpha'_{\p}\Lb 0 \Rb\,\,\,=\,\,\,0,
\eeq
in the limit of $\frac{2}{\sqrt{\lambda}}\,\ll\,1$. In the next
section we will discuss the main property of the Pomeron in N=4
SYM. We will introduce available models which incorporate those
properties. In the framework of N=4 SYM motivated models for soft
high-energy interaction, we will then confront these models with the
LHC recent data for the inclusive hadron production.  Our main
conclusion is: it is premature to claim that {\it soft} model is
unable to describe the LHC data, and we need to have precise values of
the cross-sections of diffraction processes in order to claim so. On the same footing, 
we should improve the Monte Carlo based simulation models in order to include those processes.

In the last section, we consider the high-density QCD picture which
provides an alternative description of soft high-energy
interaction. We will show that the high-density QCD scenario for
inclusive hadron production in proton-proton collisions works and gluon saturation reproduces
the LHC data for charged-hadron transverse-momentum and multiplicity distribution in
rapidity and energy \cite{LRPP}. We show that high-density QCD \cite{LRPP} predicted $7$ TeV data in $pp$ collisions \cite{CMS1}. We
firmly believe that if further experiments confirm that the gluon
saturation works it will be a major discovery. There exists some ideas
how to simulate the CGC state in nucleus collisions
but evidence for the formation of the CGC state (gluon saturation) in
proton-proton interaction will be a triumph of the high-density QCD.  The
last section is devoted to our predictions for the LHC both for
hadron-hadron ($pp$) and nucleus-nucleus ($AA$) collisions \cite{LR2}. We consider
this section as a key part of this manuscript since we believe that
comparison with our predictions will provide the first hint toward
discovery of the new phase of QCD, the CGC at the LHC.

    \section{Inclusive hadron production in soft Pomeron models}
    First, let us summarize what we have learned about soft
    high-energy interaction during the last four decades. The
    phenomenology based on the Pomerons and their interactions has been
    very successful in description of the experimental data.  On the
    other hand all attempts to build a theory of Pomeron and Reggeons
    and their interactions have failed. As it was already mentioned, at
    present we have two theoretical guides: high energy scattering in
    N=4 SYM which gives Pomeron interactions, and matching with
    perturbative QCD where we can use the BFKL Pomeron calculus to
    obtain the scattering amplitude. Based on these two guides, one can
    formulate the basic criteria for such models:

(i) Pomeron $\Delta_\pom \approx $ 0.3 and $\alpha'_\pom \approx $
0. The large intercept of the Pomeron follows from N=4 SYM and its
value about 0.3 follows from the description of the DIS data
\cite{LEPO}.

(ii) Large Good-Walker component \cite{GW} since in N=4 SYM the main contribution stems from the processes of elastic scattering and diffraction production \cite{BST,HIM}.

(iii) Small Pomeron interaction which is of the order of  $2/\sqrt{\lambda} \ll 1$ in N=4 SYM \cite{BAR1}.

(iv) Only triple Pomeron vertex is essential to provide a natural matching with perturbative QCD \cite{BAR2}.

{
At the moment we have only four models \cite{GLMM,KMR,OS,GLMLAST} on
the market that satisfy these criteria. Unfortunately, the inclusive hadron
production was predicted in only one model \cite{GLMM}
(see Ref. \cite{GLMINC} and \fig{incl}).
  
  \begin{figure}[ht]
\centerline{\epsfig{file=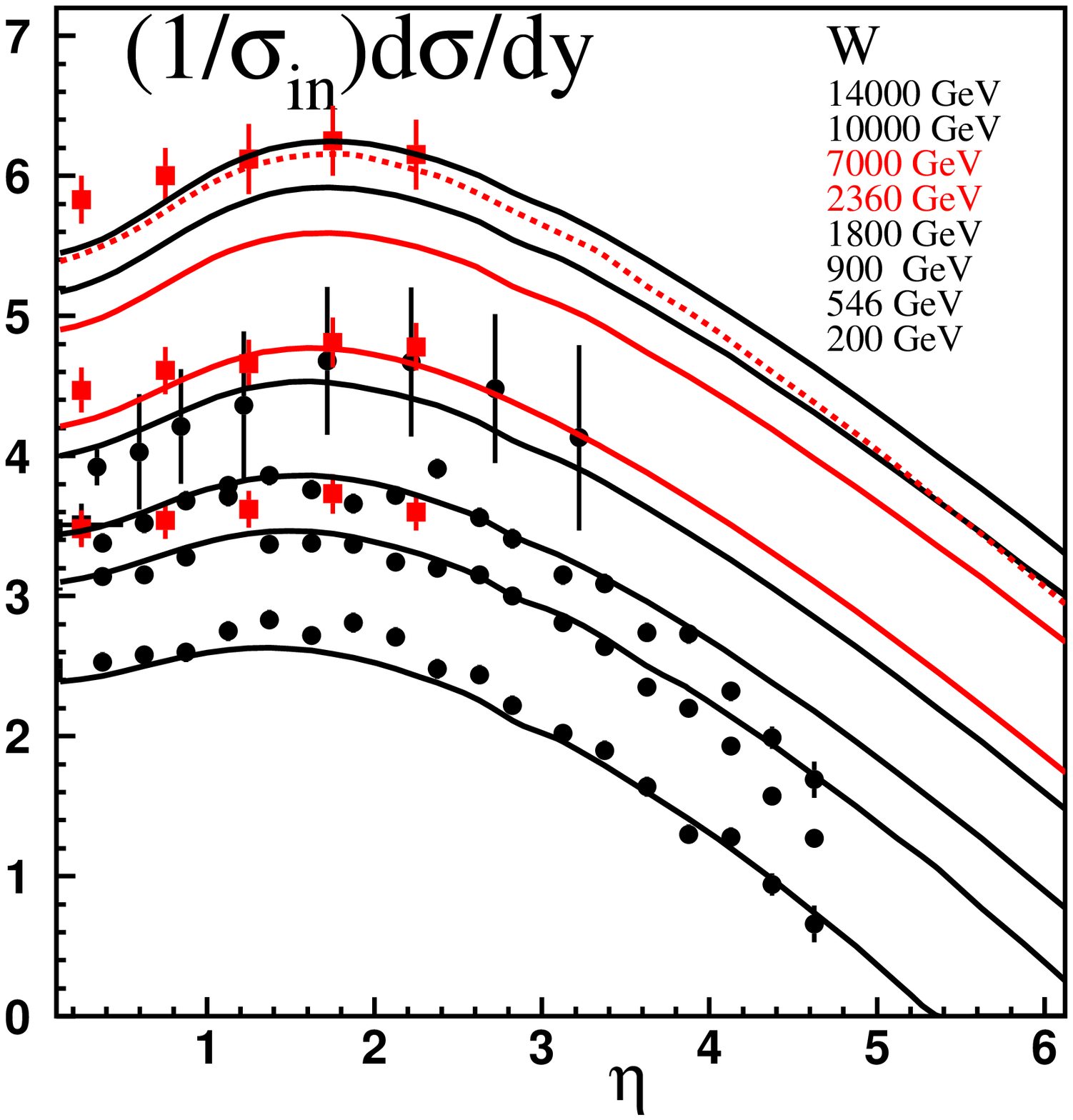,width=70mm}}
\caption{Hadron multiplicity from the soft Pomeron model \cite{GLMM} at various energy. 
 Dotted curve shows the prediction of Ref.~\cite{GLMM} divided by $\sigma_{ND}$. The experimental data are from Refs.~\cite{CMS1,AL1,CMS,particleb}. }
\label{incl}
\end{figure}

In \fig{diffeq}, we show the typical Mueller diagrams which is
used for the calculation of the inclusive hadron production. In Pomeron type model, we have two main ingredients: the full Pomeron
Green function and $\sigma_{in}$ defined in \fig{diffeq}. It turns out that the main
uncertainties stems from the value of $\sigma_{in}$. Various
predictions for $\sigma_{in}$ can be extracted from Table I.

\begin{figure}[h]
\centerline{
 \leavevmode
      \includegraphics[totalheight= 2.cm]{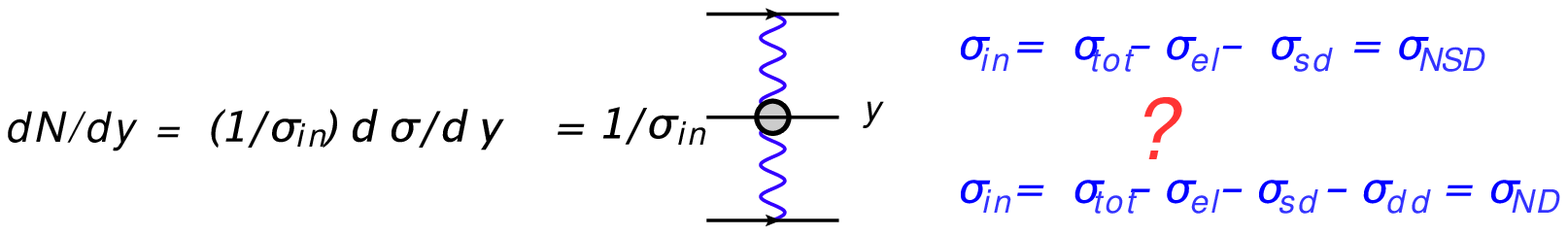}}
\caption{The Mueller diagram for the inclusive production of hadrons in Pomeron approach. The wave line denotes the full Green function of the Pomeron. $\sigma_{tot}$, $\sigma_{el}$, $\sigma_{sd}$ and $\sigma_{dd}$ denote the total, elastic, single and double diffraction 
cross-section, respectively.}
\label{diffeq}
      \end{figure}
%%%%%%%%%%%%%%%%%%%%%%%%%%%%%%%%%%%%%%%%%%%%%%%%%%%%%%%%%%%%%%%%
%{\tiny
\begin{table}[h]{
\begin{tabular}{|l|l|l|}
\hline
&\hspace{1.5cm} Tevatron (1.8 TeV)
&\hspace{1.5cm} LHC (14 TeV)
\\
& G(08)\, G(10)\, K(07)\, K(10)\, O(C)
& G(08) \, G(10) \, K(07) \, K(10) \, O(C)
 \\
\hline
$\sigma_{tot}$(mb)
& 73.29\,\,\, 74.4\,\,\,\, 74.0\,\,\,\,\, 73.9\,\,\,\,\, 73.0 
& 92.1\,\,\,\,\,\,\,\, 101\,\,\,\,\,\, 88.0\,\,\,\,\,\,\, 86.3\,\,\,\,\,\ 114.0 \\
\hline
$\sigma_{el}$(mb)
& 16.3\,\,\,\, 17.5\,\,\,\,\, 16.3\,\,\,\,\,\, 15.1\,\,\,\,\,\, 16.8 
& 20.9\,\,\,\,\,\,\,\, 26.1\,\,\,\,\,\,  20.1\,\,\,\,\,\,\, 18.1\,\,\,\,\,\ 33.0
\\
\hline
$\sigma_{sd}$(mb)
& 9.76\,\,\,\, 8.87\,\,\,\,\, 10.9\,\,\,\,\,\, 12.7\,\,\,\,\,\, 9.6 
& 11.8\,\,\,\,\,\,\,\, 10.8\,\,\,\,\,\, 13.3\,\,\,\,\,\,\, 16.1\,\,\,\,\,\ 11.0  \\
\hline
$\sigma_{dd}$(mb) 
& 5.36\,\,\,\, 3.53\,\,\,\,\, 7.2\,\,\,\,\,\,\,\, 13.3\,\,\,\,\,\, 3.93
& 6.1\,\,\,\,\,\,\,\,\,\,\, 6.5\,\,\,\,\,\,\,\, 13.4\,\,\,\,\,\,\, 12.9\,\,\,\,\,\,\,\ 4.83 \\
\hline 
${\boldmath \sigma_{NSD}}$(mb)  
& 47.2\,\,\,\, 46.9\,\,\,\,\, 46.8\,\,\,\,\,\, 43.5\,\,\,\,\,\, 47 
&61.6\,\,\,\,\,\,\,\, 64.1\,\,\,\,\,\, 54.6\,\,\,\,\,\,\, 51.2\,\,\,\,\,\,\,\ 60\\
\hline 
${\boldmath \sigma_{ND}}$(mb)  
&41.8\,\,\,\,  43\,\,\,\,\,\,\,\,\, 41.2\hspace{1.5cm} 42
&50.1\,\,\,\,\,\,\,\,  57.6\,\,\,\,\,\, 41.2 \hspace{1.45cm} 56.2\\
\hline
\end{tabular}
\caption{Comparison of various soft models: G(08) denotes the GLMM  model \cite{GLMM} in which 
one sums only enhanced diagrams. G(10) denotes the model of the Tel Aviv
group where a general class of the Pomeron diagrams is
summed \cite{GLMLAST}.  K(07) and K(10) denote two models of the Durham
group. The 2007 predictions are from Ref.~\cite{KMR} and the
preliminary 2010 predictions are taken from the talk of A. Martin at
Diffraction 2010 workshop. OS(C) is the model developed in Ref.~\cite{OS}. }
\label{t3}} 
\end{table}
%}
%%%%%%%%%%%%%%%%%%%%%%%%%%%%

  \begin{table}[h]
\begin{tabular}{|l|l|l|l|l|}
\hline
 \,\,\,\,\,\,\,\,\,\,\,\,$\sqrt{s}$  \,\, TeV
& \,\,\,\,\,\,\,\,\,\,\,\,
& \,\,\,\,\,\,\,\,\,\,\,\, Pythia6 
& \,\,\,\,\,\,\,\,\,\,\,\,  Phojet
& \,\,\,\,\,\,\,\,\,\,\,\, GLMM \\
\hline
 \,\,\,\,\,\,\,\,\,\,\,\,\,\,0.9
& \,\,\,\,\,\,\,\,\,\,\,\, $\sigma_{ND}$(mb) 
& \,\,\,\,\,\,\,\,\,\,\,\, 34.4 
& \,\,\,\,\,\,\,\,\,\,\,\, 40.0 
& \,\,\,\,\,\,\,\,\,\,\,\, 39.23  \\
\hline
 \,\,\,\,\,\,\,\,\,\,\,\, 0.9
& \,\,\,\,\,\,\,\,\,\,\,\, $\sigma_{sd}$(mb) 
& \,\,\,\,\,\,\,\,\,\,\,\, 11.7 
& \,\,\,\,\,\,\,\,\,\,\,\, 10.5 
& \,\,\,\,\,\,\,\,\,\,\,\, 8.24   \\
\hline
 \,\,\,\,\,\,\,\,\,\,\,\, 0.9
& \,\,\,\,\,\,\,\,\,\,\,\, $\sigma_{dd}$(mb) 
& \,\,\,\,\,\,\,\,\,\,\,\, 6.4 
& \,\,\,\,\,\,\,\,\,\,\,\, 3.5 
& \,\,\,\,\,\,\,\,\,\,\,\, 3.83   \\
\hline
 \,\,\,\,\,\,\,\,\,\,\,\, 7.0
& \,\,\,\,\,\,\,\,\,\,\,\, $\sigma_{ND}$(mb)
& \,\,\,\,\,\,\,\,\,\,\,\, 48.5 
& \,\,\,\,\,\,\,\,\,\,\,\, 61.6 
&
 \,\,\,\,\,\,\,\,\,\,\,\, 51.47   \\
\hline
 \,\,\,\,\,\,\,\,\,\,\,\, 7.0
& \,\,\,\,\,\,\,\,\,\,\,\, $\sigma_{sd}$(mb)
& \,\,\,\,\,\,\,\,\,\,\,\, 13.7 
& \,\,\,\,\,\,\,\,\,\,\,\, 10.7 
& \,\,\,\,\,\,\,\,\,\,\,\, 10.2 
 \\
\hline
 \,\,\,\,\,\,\,\,\,\,\,\, 7.0
& \,\,\,\,\,\,\,\,\,\,\,\, $\sigma_{dd}$(mb)
& \,\,\,\,\,\,\,\,\,\,\,\, 9.3 
& \,\,\,\,\,\,\,\,\,\,\,\, 3.9 
& \,\,\,\,\,\,\,\,\,\,\,\, 6.46   \\
\hline
\end{tabular}
\caption{Comparison of Monte Carlo simulation models with the GLMM model \cite{GLMM}.}
\end{table}
  
The curves in \fig{incl} were calculated using $\sigma_{in}
  =\sigma_{NSD}$. The dotted curve corresponds to $\sigma_{in} =
  \sigma_{ND}$. One can see that we need to understand better what
  kind of inclusive production has been measured.  The experimental
  selection of the measured events at the moment apparently depends strongly on
  the Monte Carlo simulation that has been used. Table II shows that
  there is no Monte Carlo code on the market that can consistently describe the
  diffraction production processes.

 Having these two tables in mind, we can conclude that the claim that
 soft model is not able to describe the LHC data is premature and much
 more work is needed to prove this claim.  However, the LHC recent data includes
 some interesting measuments such as the dependence of average transverse momentum of produced hadron
 on energy and multiplicity which cannot be described in
 framework of Pomeron-type models.
 
\section{High density QCD and hadron production}

The conclusion from the previous section is that one has to look
for a more adequate approach which provides a better description of the experimental
data at the LHC and will be more closely related to QCD.

Fortunately, we have such an approach on the market: high density QCD
\cite{GLR} leads to a completely different picture
of inclusive hadron production. In this approach, a system of parton
(gluons) at high energy forms a new state of matter: Colour Glass
Condensate (CGC). In the CGC picture, at high energy, the density of
partons $\rho_p$ with a typical transverse momenta less than $Q_s$
reaches a high value, $\rho_p
\propto 1/\as \,\,\gg\,\,1$ ($\as$ is the strong coupling
constant). The saturation scale $Q_s$ is a new momentum scale that
increases with energy. At high energies/small Bjorken-$x$, $ Q_s
\,\,\gg\,\,\mu$ where $\mu$ is the scale of soft
interaction. Therefore, $\as\Lb Q_s\Rb \,\,\ll\,\,1$ and this fact
allows us to treat this system on solid theoretical basis. On the
other hand, even though the strong coupling $\alpha_S$ becomes small
due to the high density of partons, saturation effects, the fields
interact strongly because of the classical coherence. This leads to the a
new regime of QCD with non-linear features which cannot be investigated
in a more traditional perturbative approach. In the framework of the CGC approach the secondary hadrons are
originated from the decay of gluon mini jets with the transverse
momentum equal to the saturation scale $Q_s(x)$. The first stage of
this process is rather under theoretical control and determines the
main characteristics of the hadron production, especially as far as
energy, rapidity and transverse momentum dependence are concerned. The
jet decay, unfortunately, could be treated mostly phenomenologically.
%However, we can hope that the phenomenological uncertainties would be
%reduced to several constants whose values will be extracted from the
%experiment.

Actually, such a scenario has passed two critical tests with the
experimental data: First, it explains the main features of hadron
multiplicity in heavy ion-ion collisions at RHIC (KLN papers \cite{KLN}); and it
gave correct predictions for the inclusive hadron production in proton-proton ($pp$) collisions
\cite{LRPP} at the LHC at $\sqrt{s}=7$ TeV \cite{CMS1}.

The inclusive mini jet cross-section in high-energy $pp$ (or $AA$) collisions can be calculated within the CGC
approach via the $k_T$ factorization \cite{KT1,KTINC,KT2} by convolution of two hadrons (or
nucleus) unintegrated gluon distributions
$\phi^{h_i(A_i)}_{G}(x_i;\vec k_T)$, depicted in \fig{dndy}-a, where
$x_{1,2}=(p_{T}/\sqrt{s})e^{\pm y}$, $p_{T}$
and $y$ are the transverse-momentum and rapidity of the produced gluon
mini jet. The relation between the unintegrated gluon density and the
colour dipole-proton (or nucleus) forward scattering amplitude
$N_{h_i(A_i)}\Lb x_i; r_T; b \Rb$ was obtained in
Ref.~\cite{KTINC} which relates the hadron production in $pp$ (or
$AA$) collisions to deep inelastic lepton-hadron
scattering (DIS) at small Bjorken-$x$ at HERA. It reads as
follows
\beq \label{M2}
\phi^{h_i(A_i)}_{G}\Lb x_i;\vec{k}_T\Rb=\frac{1}{\alpha_s} \frac{N^2_c-1}{2(2 \pi)^3 N_c}\int d^2 \vec b d^2 \vec r_T
e^{i \vec{k}_T\cdot \vec{r}_T}\nabla^2_T N^{h_i(A_i)}_{G}\Lb x_i; r_T; b\Rb,
\eeq
with notation
\beq \label{M3}
N^{h_i(A_i)}_{G}\Lb x_i; r_T; b \Rb =2 N_{h_i(A_i)}\Lb x_i; r_T; b \Rb - N^{2}_{h_i(A_i)}\Lb x_i; r_T; b \Rb,
\eeq
where $r_T$ denotes the dipole transverse size and $\vec b$ is the impact parameter of the scattering.  
Notice that the relation between the unintegrated gluon density and the
forward dipole amplitude in the $k_T$ factorization is not a simple Fourier
transformation which is commonly used in literature and also depends on
the impact-parameter. The impact-parameter dependence in these
equations is not trivial and should not be in principle assumed as an
over-all factor.
For the dipole amplitude, we use the b-CGC saturation model \cite{WAKO} which is the generalization of the
approach given in Refs. \cite{WAMOKO,IIMDIS} and effectively incorporates
all known saturation properties \cite{LRPP} driven by the small-x non-linear evolution equations
including the impact-parameter dependence of the dipole amplitude
\cite{LTSOL}. This model describes both the HERA DIS data at small-$x$
\cite{WAKO}, direct-photon production \cite{RS} and the inclusive hadron production in $pp$ collisions \cite{LRPP}. The extension of
this model for the case of nuclear target was introduced in
Ref.~\cite{LR2} which also give a good description of RHIC multiplicity data.

\begin{figure}[t]
\begin{tabular}{c c c}
       \includegraphics[width=9 cm,height=6cm] {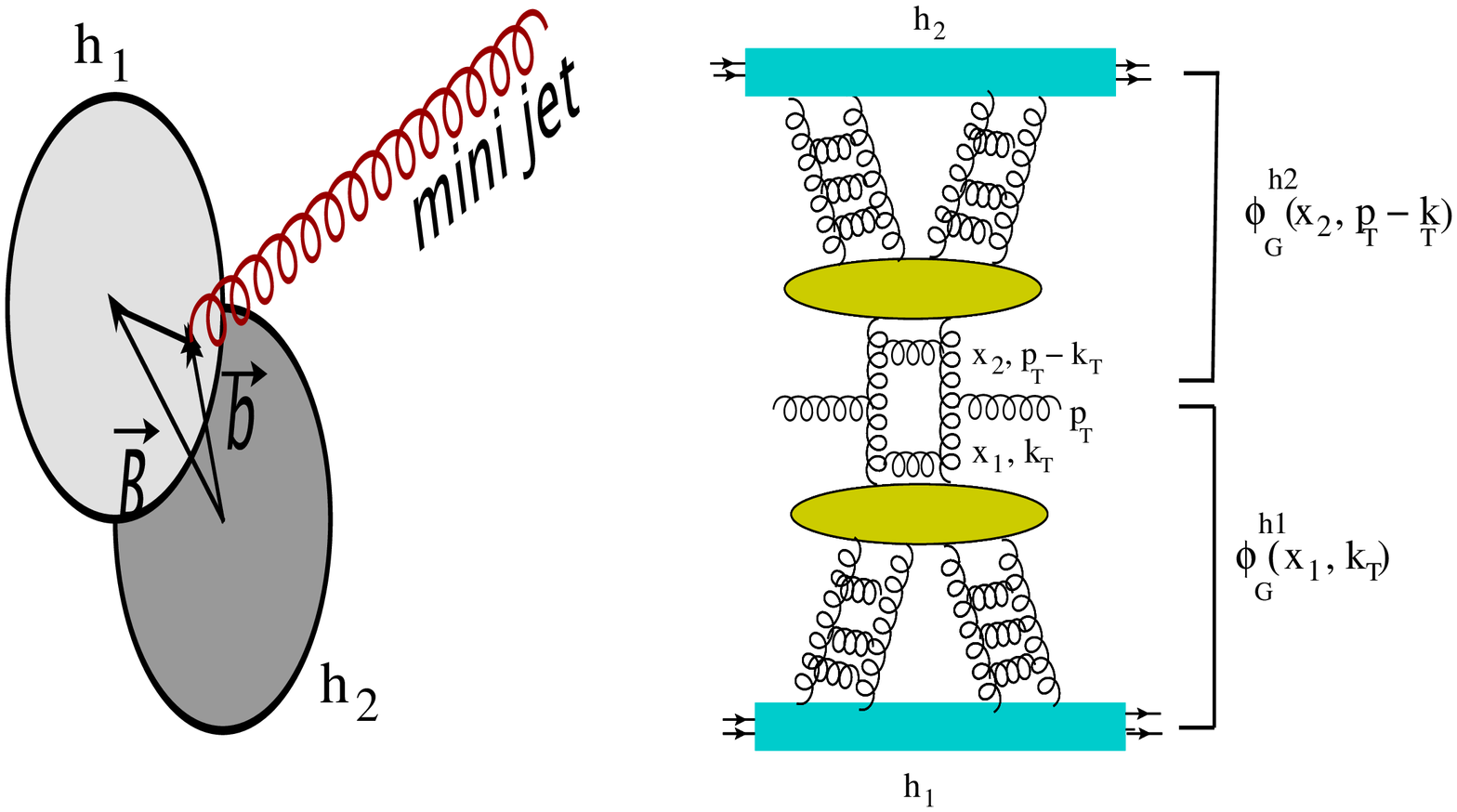} &\,\,\,\,\,\,\,\,\,&
       \includegraphics[width=8cm,height=6cm] {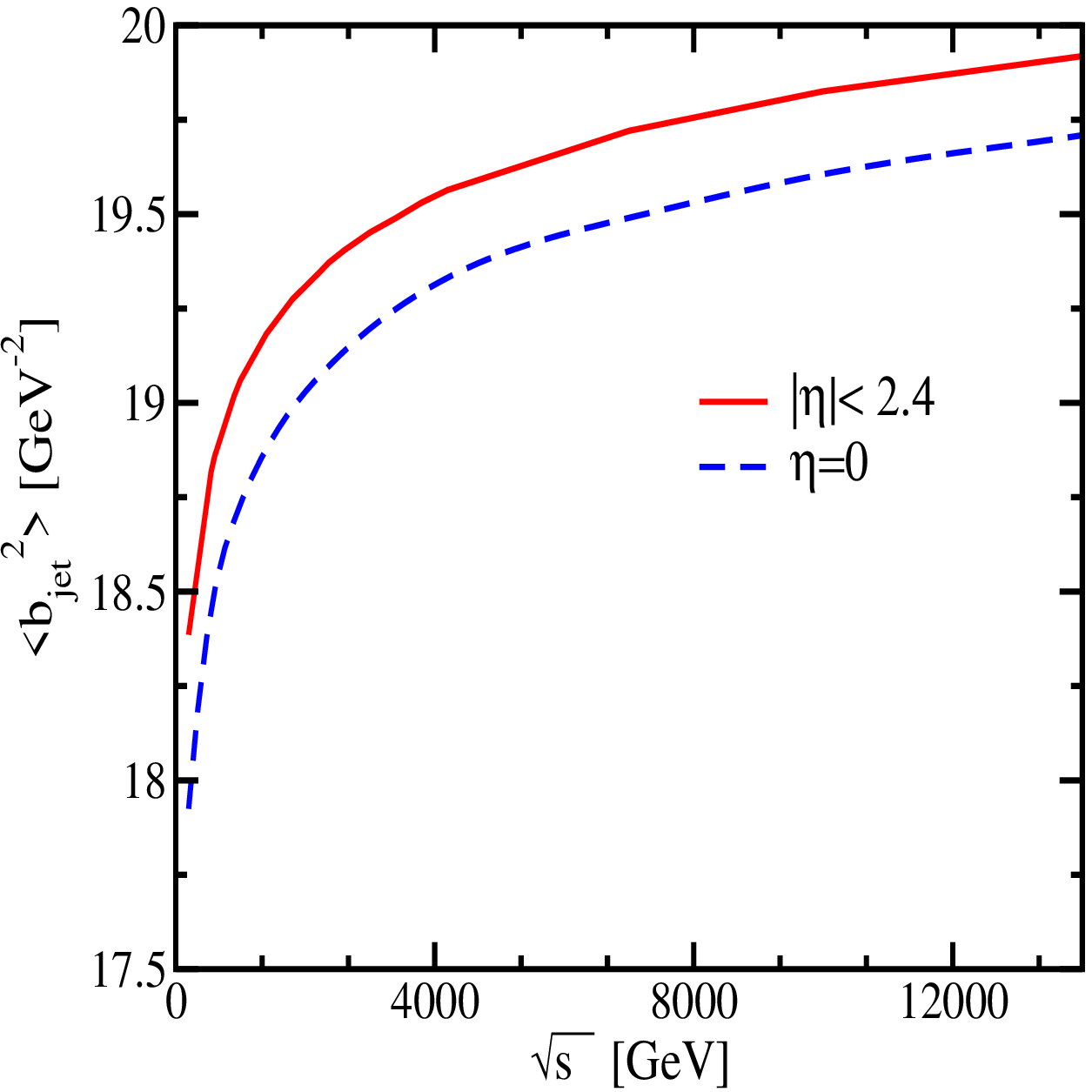}    \\
       \fig{dndy}-a & & \fig{dndy}-b\\
        \includegraphics[width=8 cm,height=6cm] {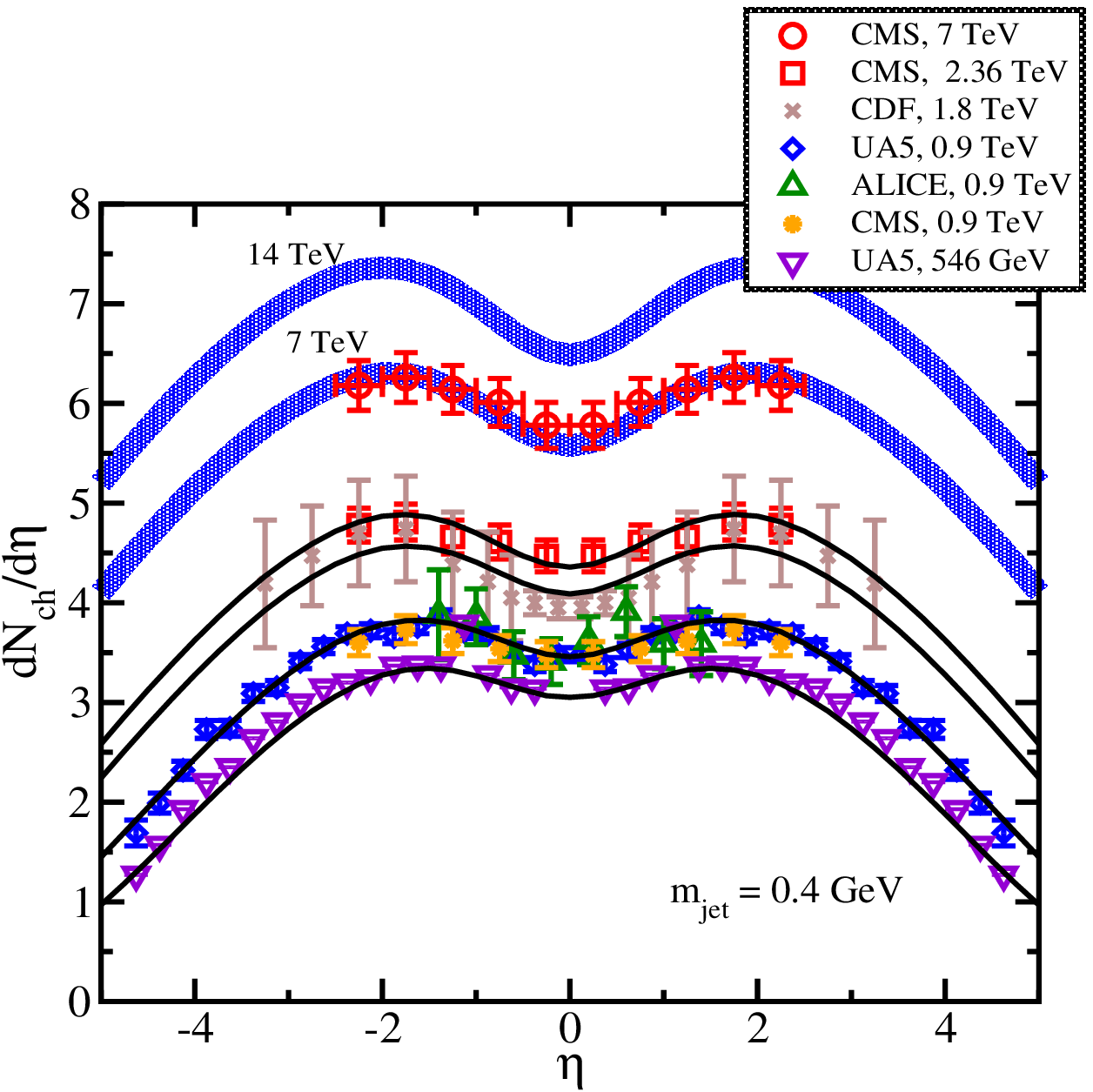} &\,\,\,\,\,\,\,\,\,&
       \includegraphics[width=8 cm,height=6cm] {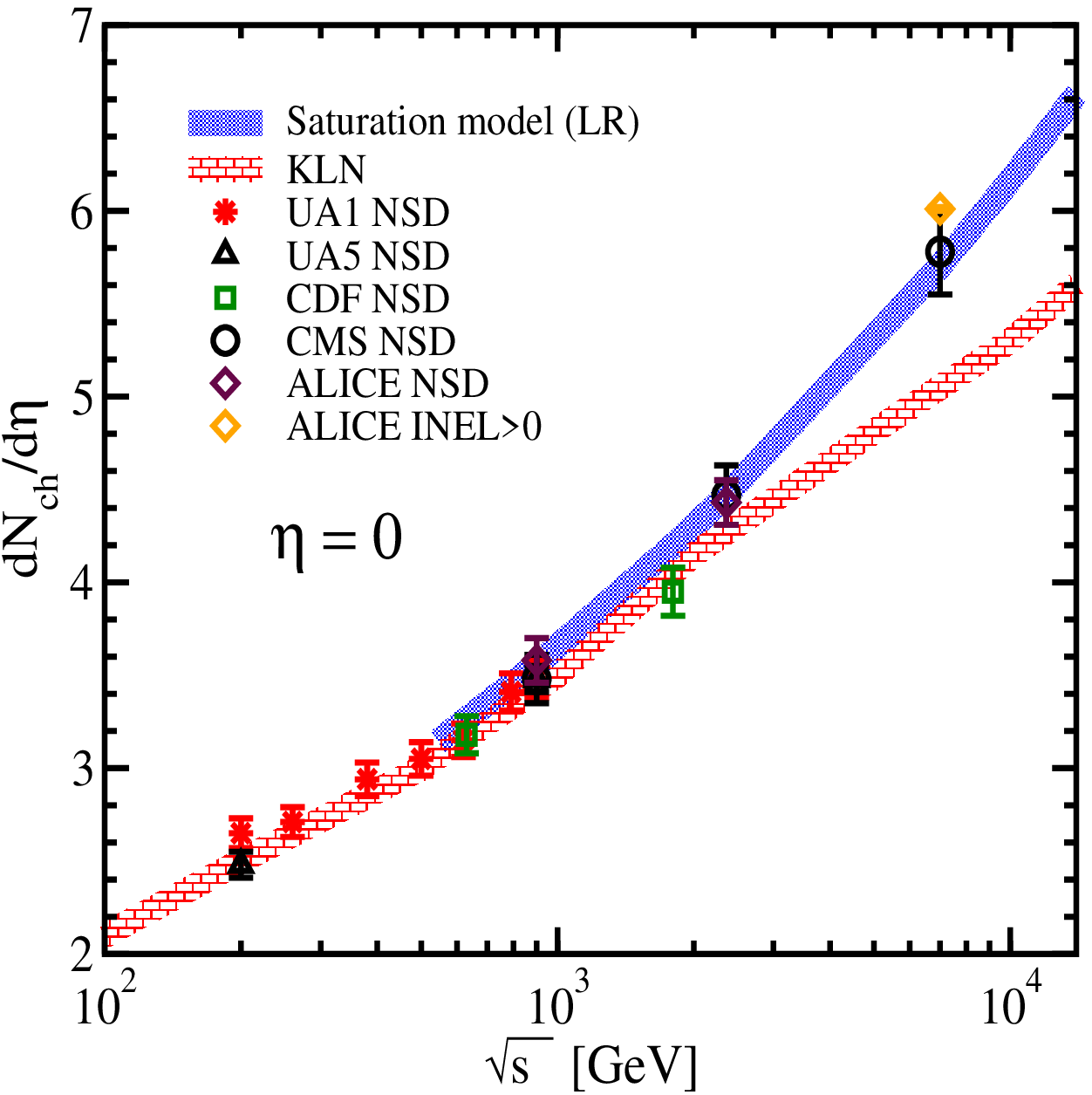}   \\
              \fig{dndy}-c & & \fig{dndy}-d\\    
                        \end{tabular}
                    \caption{a) Mini jet production in hadron-hadron collisions  
in the transverse plane within the $k_T$ factorization scheme. The impact-parameter between two hadrons is $\vec B$. 
b) shows the average impact parameter of the produced mini jet $\langle b^2_{jet} \rangle$ as a function of energy within two rapidity bins.
c)  The comparison with the experimental data and
       prediction for $dN_{ch} / d y$. The curves
       are normalized by data at $\sqrt{s} = 546\,\text{GeV}$ \cite{LRPP}. d)  Energy dependence of the charged hadrons
       multiplicity in the central region of rapidity $\eta=0$ in $pp$
       collisions. The theoretical curve (Saturation model LR) is our prediction coming from
       the saturation model \cite{LRPP}. The band indicates
       about $2\%$ theoretical error. The total theoretical uncertainties is less $6\%$ at high energies. 
       We also show the KLN prediction \cite{KLN} with the same error band as ours. Notice that in c panel we have taken a fixed mini jet mass 
$m_{jet}=0.4$ GeV for all energies while in $d$ panel uncertainties due to the assumption of a fixed energy-independent mini jet mas was included in the band. The experimental data are from
       Refs.~\cite{CMS1,AL1,CMS,particleb,ua1}. The
       experimental error bars indicate systematic uncertainties. }
  \label{dndy}   
         \end{figure}

The main contribution of the $k_T$
factorization in the multiplicity comes from $p_T< 2$ GeV. For such a
kinematic region at very low-$p_T$, we rely on the Local Parton-Hadron
Duality principle \cite{LPHD}, namely we assume that the hadronization
is a soft process and cannot change the direction of the emitted
radiation. This works perfectly in $e^+ e^- $ annihilation into hadrons
\cite{LPHD,LPHD1} and we believe that this is more preferable than to
deal with the fragmentation functions for which we have no
theoretical justifications at low $p_T$. Hence, the form of the rapidity
distribution of the mini jet and the produced hadron Eq.~(\ref{PO9})
are different only with a numerical factor $C$, and the transverse
momentum of jet and the produced hadron are related with a factor
$\langle z\rangle$.  In the spirit of the geometrical-scaling property
of the scattering amplitude, we obtain the charged-particle
multiplicity distribution at a fixed centrality but various energies
from the corresponding mini jet cross-section divided by the average
area of interaction $\sigma_{s}\propto \pi \Big \langle \vec b^2_{jet}
\Big\rangle$, see \fig{dndy}-b. The $p_T$ spectrum of the produced hadron can be then related to the
cross-section of the mini jet production in the following way: 
\beq \label{PO9}
\frac{d N_{\mbox{hadron}}}{d^2 p_T}\,\,=C\int d \eta\,h[\eta]\frac{1}{\sigma_{s}}\, \,\frac{d \sigma^{\text{mini jet}}}{d \eta \,d^2 p_{\mbox{jet},T}}\left[\mbox{with}\,\, p_{ \mbox{jet}, T} \,= \,p_T/\langle z\rangle \right].
\eeq
Notice that $k_T$ factorization has infrared divergence. By
introducing a new parameter $m_{eff}$ as mini jet mass which mimics the
pre-hadronization effect, one can also regularize the cross-section. Therefore,
we have only two unknown parameters in our model, the overall factor $C$
and the mini jet mass $m_{jet}$ which are fixed at lower energy. Then
our results at higher energies and rapidities can be considered as
free-parameter predictions.

\fig{dndy} shows our description of the existing experimental data
 and predictions for higher energies. In this figure, we show for the
 first time the comparison of our prediction \cite{LRPP} with $7$ TeV
 $pp$ data \cite{CMS1}. One can see that the agreement is striking. It should be
 stressed that our approach gives a quite different result for the LHC
 energies in comparison with the Kharzeev-Levin-Nardi (KLN) approach \cite{KLN}
 both for $pp$ and $AA$ collisions. The main differences stem from the
 fact that: 1) we used a saturation model that describes the HERA data
 at small-$x$ and has different energy dependence and value for the
 saturation scale. 2) We used a correct relation between the
 unintegrated gluon-density and the forward dipole-nucleon amplitude
 in the $k_T$ factorization, namely Eqs.~(\ref{M2},\ref{M3}). 3) In
 contrast to the KLN, we kept explicitly the impact-parameter
 dependence of the formulation and did not assume that it is trivially
 factorizable as a normalization factor. 4) The relative increase of
 $\sigma_s$ was calculated in our approach while in the KLN approach
 was taken from soft high-energy interactions which is alien to the
 saturation approach. In both approaches, lower energy data was used
 to fix the overall normalization factor. However, as can be seen from
 Figs.~\ref{dndy}-d, \ref{fa}-b, we expect that the discrepancies between our
 predictions and the KLN to be more pronounced at higher energies
 (even more in AA collisions).
\begin{figure}[t]
\begin{tabular}{c c c}
       \includegraphics[width=8 cm,height=6cm] {plot-avp2.eps} &\,\,\,\,\,\,\,\,\,&
       \includegraphics[width=8cm,height=6cm] {plot-p2.eps}    \\
       \fig{dndpt}-a & & \fig{dndpt}-b\\
        \includegraphics[width=8 cm,height=6cm] {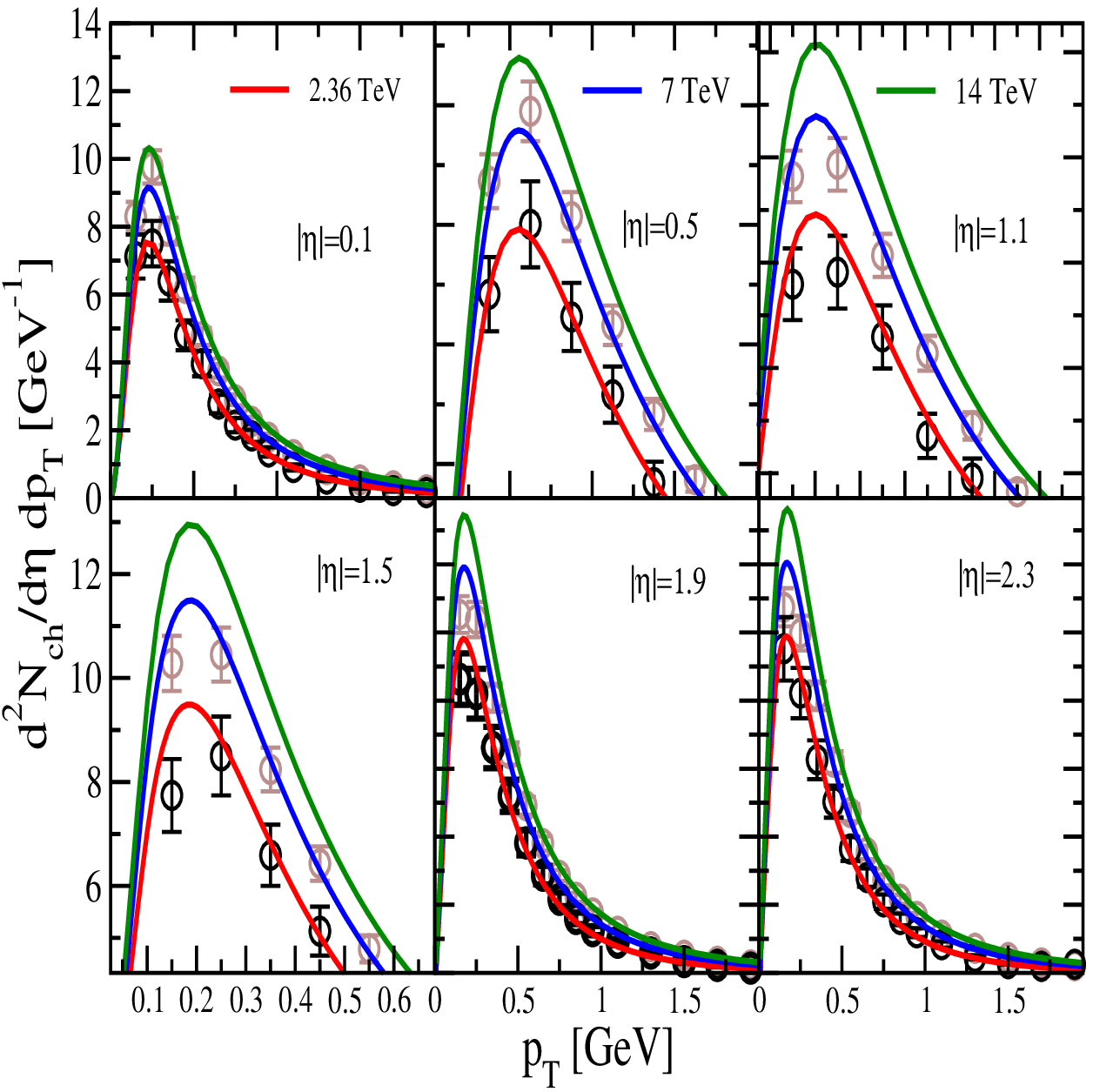} &\,\,\,\,\,\,\,\,\,&
          \includegraphics[width=8 cm,height=6cm] {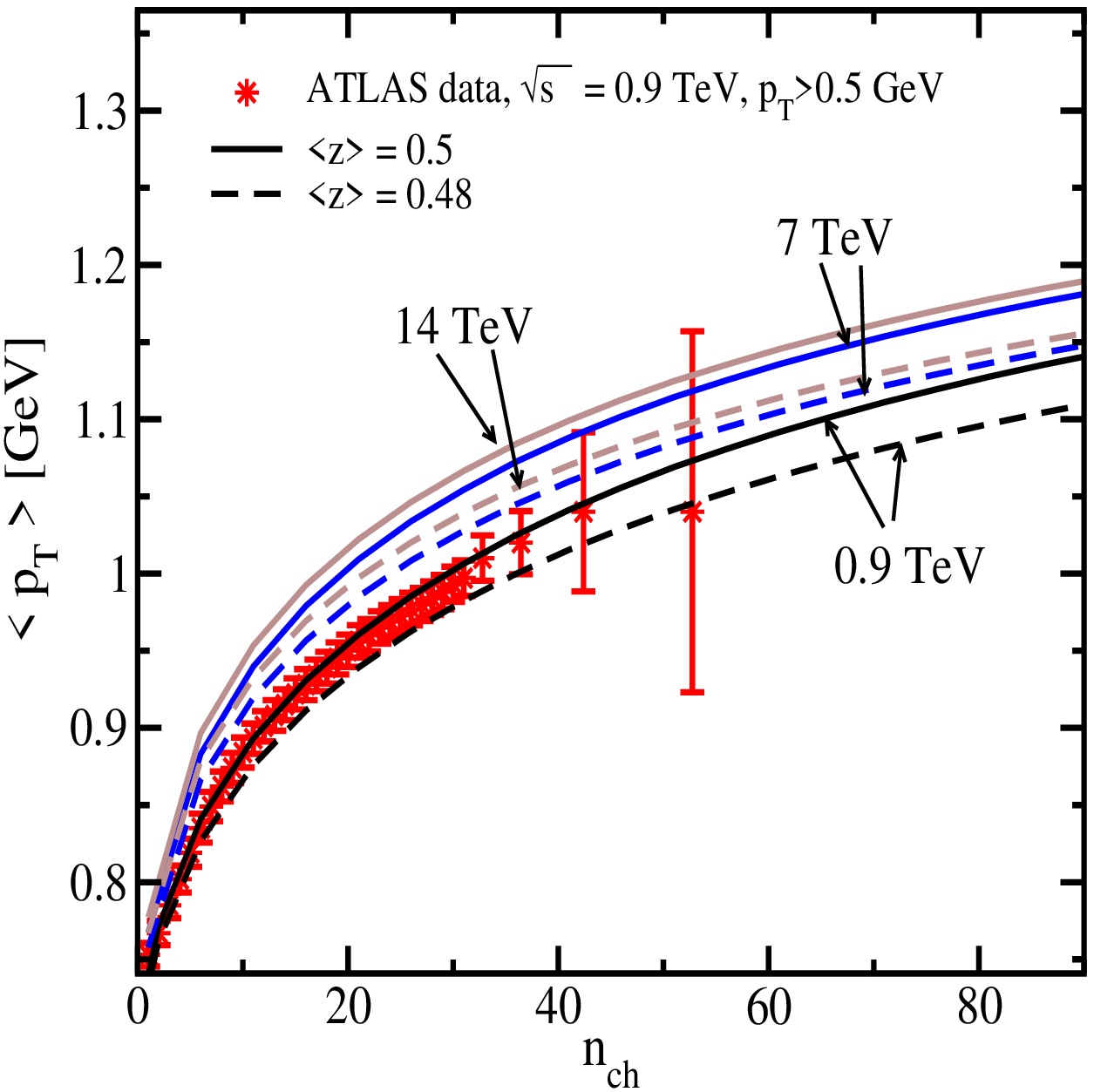}  \\
              \fig{dndpt}-c & & \fig{dndpt}-d\\    
                        \end{tabular}
                    \caption{a) The energy dependence of the average transverse momentum of charged hadrons.
b,c)The differential yield of charged hadrons. The LHC experimental data are from the CMS collaboration \cite{CMS1,CMS}.
d) The average transverse momentum of charged hadrons as a function of the number of charged particles for 
events with $n_{ch}\geq 1$ within the kinematic range
$p_T>500~\text{MeV}$. The experimental data are from ATLAS for
$\sqrt{s}=0.9$ TeV and $|\eta|<2.5$ \cite{ATLAS}. The theoretical curves was
obtained for $|\eta|=0$ and with the same kinematic constraint $p_T>500~\text{MeV}$ at various energies for two value of $\langle z\rangle =0.48, 0.5$ corresponding to the dashed and the solid lines, respectively.
 The mini jet mass is taken $m_{jet}=0.4$ GeV in all plots. The normalization is the same as in \fig{dndy}.               
               }
               
                   \label{dndpt} 
         \end{figure}   

In order to obtain the average transverse momentum of charge hadrons,
we need also to know the value of the average fraction of energy of
mini jets carried by the hadrons $\langle z\rangle$. It is seen from
Fig.~ \ref{dndpt}-a that an average value of $\langle z
\rangle=0.48\div 0.5$ is remarkably able to describe the average
transverse momentum of charge hadrons in a wide range of energies.  In
order to further test the validity of the value $\langle z \rangle$
for the mini jets, we show in Figs.~\ref{dndpt}-b, \ref{dndpt}-c our
predictions for the differential yield of charged hadrons in the range
$|\eta|<2.4$ and at various $|\eta|$ bins for $\sqrt{s}=2.36, 7$ and $14$ 
TeV. The experimental data are recently reported from the CMS
collaboration \cite{CMS1,CMS}. It is seen that our predictions is in quite good
agreement with experimental data at $7$ TeV.  We recall again that the
pre-factor in Eq.~(\ref{PO9}) is the same as what we already fixed
with experimental multiplicity data at low-energy $\sqrt{s}=546$ GeV
at $\eta=0$ in \fig{dndy}-c. Therefore, we have no free parameters in obtaining the
theoretical curves in Fig.~\ref{dndpt}. The fact that our model
reasonably works at low $p_T$ is due to the fact that the saturation
scale is rather large at low $p_T$. In our formulation, we predicted
that the differential yield of charged hadrons has a peak at low
$p_T$. The position of the peak is approximately at $p_T\simeq m_{jet}
\langle z\rangle $ \cite{LRPP}. The experimental data at $7$ TeV shown in
Figs.~\ref{dndpt}-d indeed confirmed this prediction. 
In the CGC scenario, the gluon saturation scale is proportional to the
density of partons. The parton density is proportional to the
multiplicity and, therefore, one can relate the saturation momentum in
the event with the multiplicity of the hadrons $n$. In
Fig.~\ref{dndpt}-d, we show the average transverse momentum of charged
hadrons as a function of the number of charged particles for events
within the kinematic range $p_T>500~\text{MeV}$ at various
energies. The experimental data are from ATLAS for $\sqrt{s}=0.9$ TeV
\cite{ATLAS}. Our prediction seems also to be in a good agreement with preliminary $7$ TeV data from the ATLAS collaboration (not shown in the figure).

Finally, in Fig.~\ref{fa} we show our prediction for Pb+Pb collisions at the LHC \cite{LR2}.
Notice that again similar to the case of $pp$ collisions, we have here only two free parameters,
normalization factor $C$ and the mini jet mass which are fixed at RHIC energy $\sqrt{s}=200$ GeV for $0-6\%$ centrality. 
Therefore, at lower/higher energies than $\sqrt{s}=200$ GeV (for various centrality/rapidities) we do not have any free parameter. 
In Fig.~\ref{fa}-a, we show our predictions at lower RHIC
energies $\sqrt{s}=19.6$ and $130$ GeV in Au-Au collisions, and also
for the LHC energies $\sqrt{s}=2.75$ and $5.5$ TeV in Pb-Pb collisions for
$0-6\%$ centrality bin. In Fig.~\ref{fa}-b, we show the energy dependence of $dN_{pp}/d\eta$,
$dN_{AA}/d\eta$ and $(2/N_{par})dN_{AA}/d\eta$ at midrapidity $\eta=0$
for central collisions (where $N_{par}$ denotes the number of
participant for a given centrality).

To conclude, we showed that the CGC gives very good descriptions of
the first data from the LHC for the inclusive charged-hadron
production in proton-proton collisions, the deep inelastic scattering
at HERA at small Bjorken-$x$, and the hadron multiplicities in
nucleus-nucleus collisions at RHIC. We believe that our predictions for
nucleus-nucleus collisions at the LHC will be a crucial test of the CGC
approach.

\begin{figure}[t]
\begin{tabular}{c c c}
       \includegraphics[width=7 cm,height=6cm] {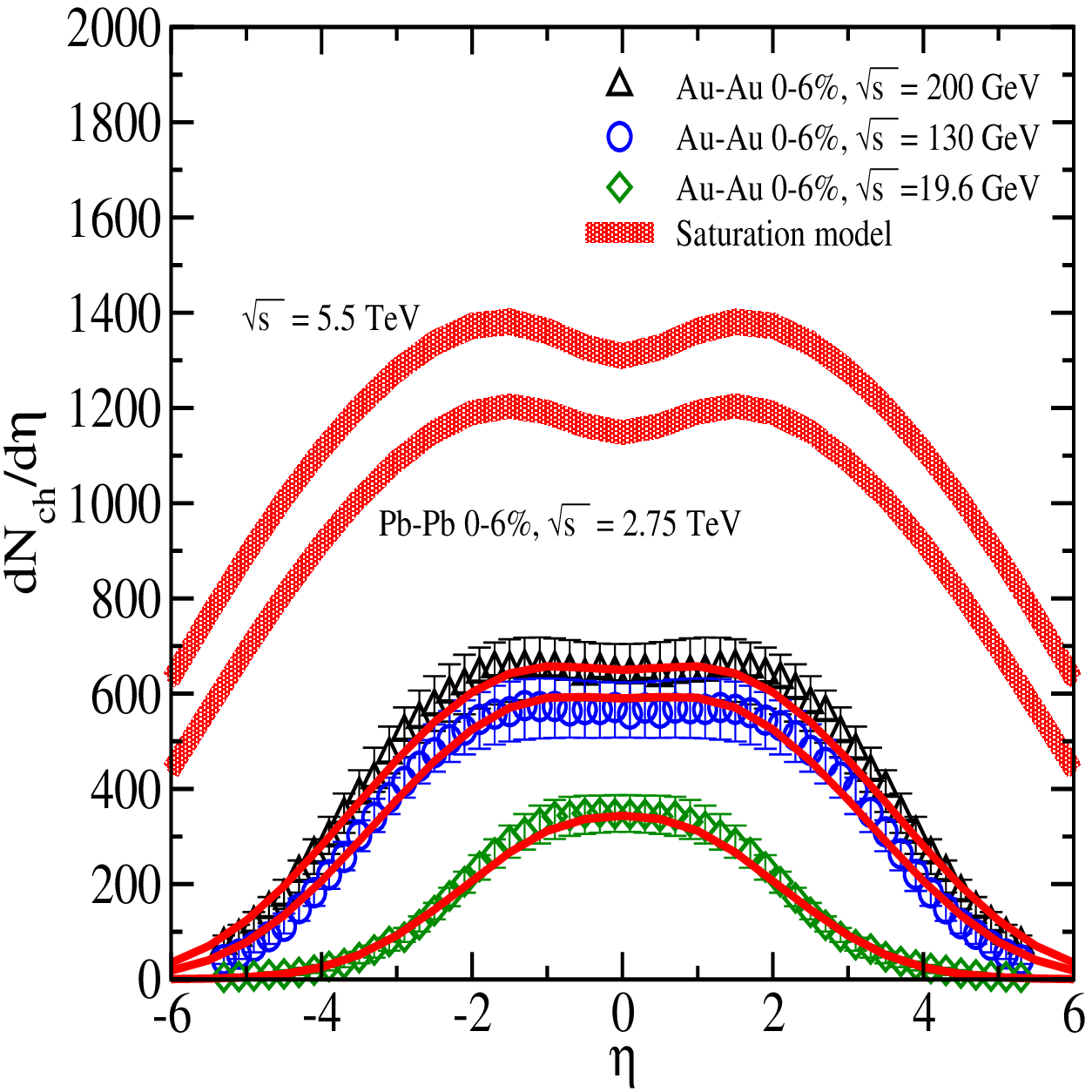} &\,\,\,\,\,\,\,\,\,&
       \includegraphics[width=7cm,height=6cm] {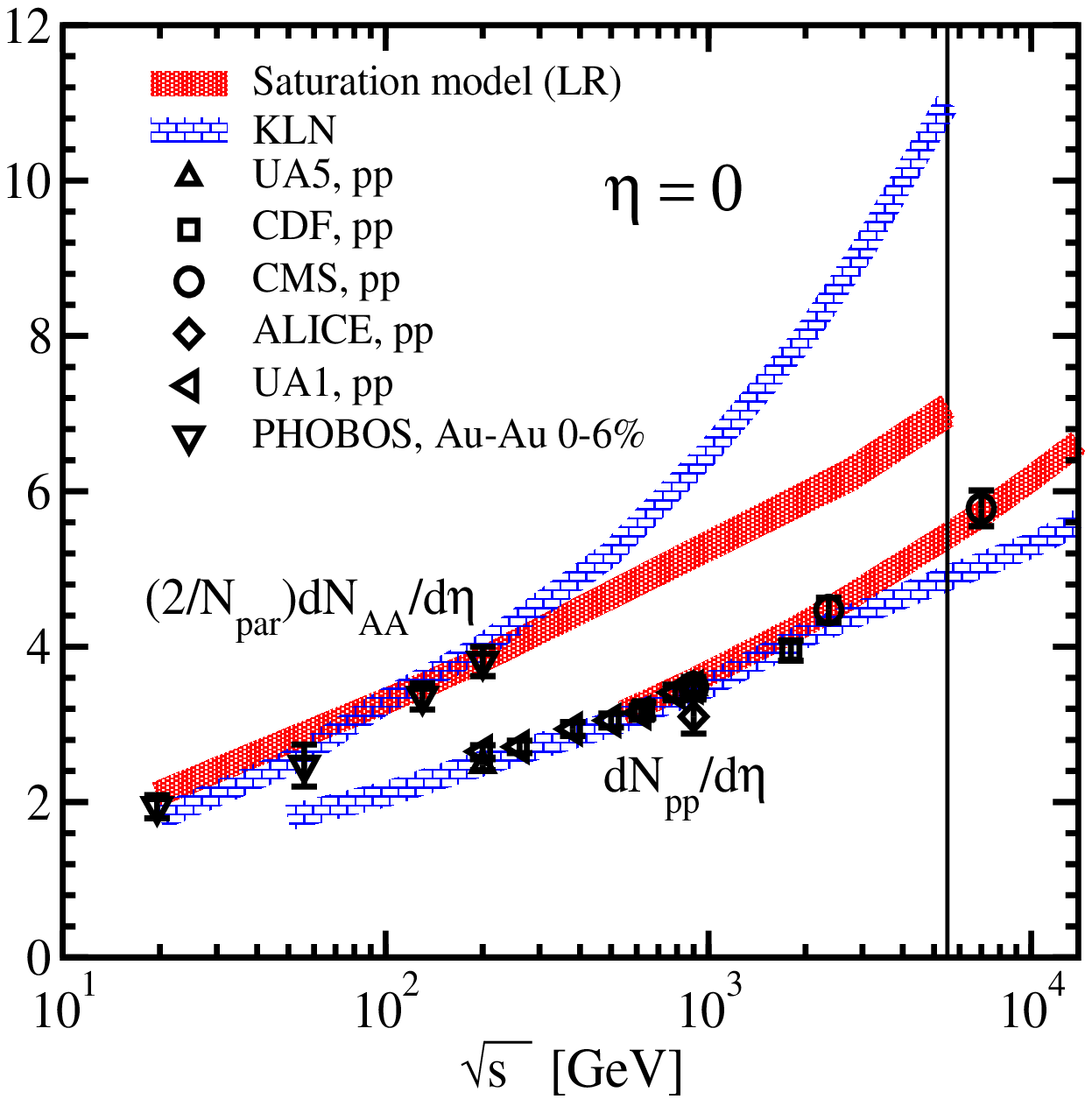}    \\
       \fig{fa}-a & & \fig{fa}-b\\
\end{tabular}
                    \caption{a) Pseudo-rapidity distribution of charged particles
             produced in Au-Au and Pb-Pb central $0-6\%$ collisions at RHIC
             and the LHC energies. b) Energy dependence of the charged hadrons multiplicity at midrapidity $\eta=0$ in central collisions in $pp$ and $AA$ collisions.  The theoretical curve Saturation model (LR) is our prediction. The band indicates less than $3\%$ theoretical errors. The total theoretical uncertainties is less than $7\%$. The experimental data are from \cite{CMS1,AL1,CMS,particleb,ua1,rhic1}. The plots are taken from Ref.~\cite{LR2}.       
               }
                                \label{fa}   
         \end{figure}

\section{acknowledgments}
E. L. would like to thank A. Gotsman, U. Maor and J. Miller for
fruitful collaboration. This work was supported in part by Fondecyt
grants 1090312 and 1100648.

\end{document}